# Towards a Capability Assessment Model for the Comprehension and Adoption of AI in Organisations

Authors

Tom Butler, Angelina Espinoza-Limón[a] and Selja Seppälä

BIS Department, University College Cork, Ireland

**Abstract**

*The comprehension and adoption of Artificial Intelligence (AI) are beset with practical and ethical problems. This article presents a 5-level AI Capability Assessment Model (AI-CAM) and a related AI Capabilities Matrix (AI-CM) to assist practitioners in AI comprehension and adoption. These practical tools were developed with business executives, technologists, and other organisational stakeholders in mind. They are founded on a comprehensive conception of AI compared to those in other AI adoption models and are also open-source artefacts. Thus, the AI-CAM and AI-CM present an accessible resource to help inform organisational decision-makers on the capability requirements for (1) AI-based data analytics use cases based on machine learning technologies; (2) Knowledge representation to engineer and represent data, information and knowledge using semantic technologies; and (3) AI-based solutions that seek to emulate human reasoning and decision-making. The AI-CAM covers the core capability dimensions (business, data, technology, organisation, AI skills, risks, and ethical considerations) required at the five capability maturity levels to achieve optimal use of AI in organisations. The AI-CM details the related individual and team-level capabilities needed to reach each level in organisational AI capability; it, therefore, extends and enriches existing perspectives by introducing knowledge and skills requirements at all levels of an organisation. It posits three levels of AI proficiency: (1) Basic, for operational users who interact with AI and participate in AI adoption; (2) Advanced, for professionals who are charged with comprehending AI and developing related business models and strategies; and (3) Expert, for computer engineers, data scientists, and knowledge engineers participating in the design and implementation of AI-based technologies to support business use cases. In conclusion, the AI-CAM and AI-CM present a valuable resource for practitioners, businesses, and technologists, looking to innovate using AI technologies and maximise the return to their organisations.[b]*

## 1  Introduction

Once the domain of science fiction, Artificial Intelligence (AI) now permeates society: It powers applications in our phones, offices, the factories that manufacture our goods, and increasingly the digitalisation of business services. The current ubiquity of AI was made possible by (a) The digital transformation of business and society; and (b) recent advances in data science, machine learning, and natural language processing that permit "pattern recognition and smarter ways of automating data usage". [1]

---

[a] Contact Email: aespinoza -at- xanum.uam.mx
[b] The AI-CAM and AI-CM are also available on GitHub: https://github.com/tgbutler/AI-Capability-Assessment-Model-.

The huge increase in computer processing power at all endpoints, from smartphones to cloud computing servers, saw BigTech firms invest heavily in AI R&D to advance its application.[2] Take, for example: (a) in finance, FinTech and RegTech firms emerged with powerful AI-based solutions for smart fraud detection[3,4] and financial regulation risk management;[5] (b) in health-care, AI supports patient diagnostic;[6,7,8] (c) in energy, AI is used for smart consumption;[9,10] and, (d) in retailing, for identifying customer consumption patterns.[11,12]

AI comprehension and adoption capabilities and business use cases differ widely across business domains and industry sectors. Market research illustrates that 80% of large organisations plan to adopt or have adopted some form of AI, whereas only 8% of organisations have adopted AI to support core activities.[13,14] Others have focused on single-pilot projects to assess business use cases.[15]

Research by MIT and The Boston Consulting Group[16] reveals that managers in 70% of 3000 firms studied say they understand the role of AI in generating business value. In comparison, 59% of those firms stated they had an AI strategy in their organisation, and 57% were already implementing AI-based solutions. These figures show a significant increase in AI adoption since their previous report in 2017.[17]

However, our unpublished research indicates that many firms engage in AI-washing, so the actual use of AI in business may be much less than admitted. Then, it appears that AI means different things to different people, indicating a comprehension problem. Most business use cases of AI involve sophisticated digital pattern matching, not reasoning and complex problem-solving. Classifying all such applications like AI, therefore, requires clarification.

The route to successful innovation using AI technologies is complex. Swanson and Ramillier state that "Even if an organization provisionally adopts the innovation, charting a successful course for implementation demands continued vigilance in maintaining and updating its conceptual framework, in part through reflection on its own practical experience, but also by monitoring closely the further evolution of the technology, its envisioned applications, and its marketplace. All this creates in the adopter organization an intensive need for information".[18] They argue for the development by senior management of an 'organising vision' whose purpose is to reveal organisational opportunities for technology exploitation through (1) the 'interpretation' of the technology's capabilities to achieve an understanding of how to innovate using it; (2) 'legitimation' of the need for its adoption and role in enabling or supporting business processes and decision making; and (3) 'mobilisation' of resources, financial, technical and human, internal and external, to help the adoption, implementation and use of the technology. These are important considerations and inform this paper's approach and contribution.

In business contexts, AI comprehension and adoption is the remit of professionals who comprehend, plan and adopt AI to support well-defined business use cases. However, a question arises as to whether business and technology professionals have the required AI knowledge and skills (i.e., capabilities) to address this challenge.

According to a 2017 report by *MIT Sloan Management Review*,[19] one of the main barriers to AI introduction in organisations relates to the paucity of business and technology capabilities. The report identifies specific challenges for executives to (1) Understand AI features and functions, (2) Understand suitable business use cases for the application of AI, and (3) Develop AI-based competitive strategies. It also shows a disparity in AI knowledge and skill levels, particularly extant understandings of the full spectrum of AI technologies and how these can be applied in business use cases to yield a return on investment (ROI).[20]

In response to the AI business imperative, several problems confront firms: (1) Should they upskill employees? Or (2) employ or contract in AI professionals or capabilities? Or (3) outsource AI projects to professional services firms of AI vendors? This means, business leaders, professionals and technologists need to properly comprehend AI to ensure an optimal fit between AI technologies and business uses cases. Thus, business professionals and technologists in firms need the capabilities to: (1) Design AI adoption strategies; and (2) Enable employees to acquire AI skills commensurate with their role and level of involvement in AI projects (e.g., managerial and technical). This capabilities requirement applies whether solutions are provided in-house or are outsourced.

The challenge for firms is: how do they guide AI knowledge and skills acquisition in an organisational context? Understanding digital technology is a major problem for business executives—for example, many are not aware of the difficulties in sourcing and deploying AI expertise.[21] Furthermore, they are unaware of the range of AI knowledge and skills required across several disciplines, from data and computer science to knowledge representation, system engineering, and business subject matter expert (SME) expertise. However, to understand why AI capabilities are multidisciplinary, it is necessary to understand what Artificial Intelligence is—Section 2 sets the stage by addressing this fundamental issue.

This article's contribution is a Capability Assessment Model for AI adoption in organisations: This specifies the capabilities required to achieve desired organisational outcomes. Ceteris paribus, the higher the level of capabilities demonstrated across the various dimensions, the greater the success in AI innovation. Therefore, an organisation can determine if a capabilities gap exists and use the model and the related AI Capabilities Matrix to close the gap in line with achieving the strategic and operational benefits of AI adoption.

The remainder of this paper is structured as follows: As indicated, Section 2 presents contrasting definitions of AI to inform the development and understanding of the AI-CAM and the AI-CM. Section 3 presents the AI Capability Assessment Model. Section 4 presents the AI Capabilities Matrix, which indicates the knowledge and skills that need to be acquired by the organisation's staff according to required proficiency levels and the organisation's AI capability maturity level (1-5). Section 5 briefly discusses the contribution of this paper and offers some concluding thoughts.

## 2   Understanding AI and its Capabilities

There are different conceptions and understandings of Artificial Intelligence and its capabilities.[22] The Alan Turing Institute[23] defines AI as "the science of making computers do things that require intelligence when done by humans ".[24] A broader definition is that "AI is the theory and development of computer systems able to perform tasks normally requiring human intelligence, such as visual perception, speech recognition, decision-making, and translation between languages" .[25] Another definition is that "AI is a set of tools and technologies that has the ability to augment and enhance organisational performance. This is achieved by creating artificial systems to solve complex environmental problems, with Intelligence being the simulation of human-level intelligence" .[26]

AI is also considered a general-purpose technology (GPT) with a unique learning capability that provides organisations with potentials for wide-ranging improvements, as well as entirely new business opportunities.[27] To achieve AI's attributed outcomes involves applying concepts and techniques from machine learning, deep learning, natural language processing, knowledge representation, expert systems, and robotics.[28,29]

Thus, the following all-inclusive definition is adopted: "AI refers to multiple technologies that can be combined in different ways to:" sense, comprehend, and act, with "the ability to learn from experience and adapt over time."[30] *Sense* refers to the capability to perceive the world around the system (images, sound and speech) through technologies such as computer vision and audio processing; *Comprehend* refers to the capability to analyse and understand the data and information through inferencing; *Act* refers to a capability to take action through decision-making, choosing among competing courses of action, and so on. In the round, this is considered as weak or narrow AI.[31]

Narrow AI is all that is currently possible, at least commercially. As a computing paradigm, it has three important foci:[32]

1. **Perceptual computing**: This involves capturing and interpreting signals from the environment. It is related to pattern recognition and classification for numerical data and digitised visual, audio or analogue text. Perceptual computing includes a family of machine learning (ML), including deep learning (DL), algorithms requiring more or less human intervention for their implementation, such as supervised (when all data is labelled to train algorithms), semi-supervised (when data is partially labelled), and unsupervised (algorithms need to discover the data structure without human supervision) learning. Some ML and DL applications focus on natural language processing (NLP) and image recognition (vision computing).

2. **Semantic computing**: This involves capturing the meaning of concepts and the relations between them. Thus, it is the knowledge representation of the business entities expressed as data. Semantic computing is mostly used to represent and organise data into a knowledge base (KB), which is a graph of connected nodes through links, where the nodes are the business entities and the links, the relationships between the entities. A KB also allows powerful inference if its underlying model (e.g., an ontology) is defined in an ontology language.

3. **Cognitive computing**: This involves building upon and integrating perceptual and semantic computing to perform complex computing tasks based on algorithms that can learn, extrapolate, and explain how and why particular decisions/actions were taken.

These three AI categories are used to drive the development of the capability assessment model presented in the next section.

## 3  Towards an AI Capability Assessment Model (AI-CAM)

For AI adoption to be successful in organisations, senior decision-makers need to have an organising vision that informs actors on the requisite structures and processes required to ensure the successful comprehension, adoption, implementation and use of digital technologies,[33] including AI. Logically, this means that an organisation needs to have the requisite capabilities, or the capability to acquire them, through organisational learning. As this is an incremental process that requires elaboration and guidance, that is the purpose of this paper.

The AI Capability Assessment Model (Table 3) presented herein provides guidance on the core capability dimensions (i.e., business, data, technology, organisation, AI skills, risks, and ethical considerations) required at different stages and levels in AI comprehension and adoption. Thus, an organisation will be able to align its goals and objectives and the capabilities required at each level.

The AI-CAM is inspired by two well-known maturity models: The Capability Maturity Model Integration[c] (CMMI) V2.0 and the Maturity Model for the Enterprise Knowledge Graph[d] (EKG/MM). The CMMI is a model that focuses on maturing the software development process in organisations. The EKG/MM is an industry standard model definition of the capabilities that an organisation needs for creating an enterprise knowledge graph. The latter model focuses on certain aspects of the adoption of AI's semantic computing approach; however, the AI-CAM focuses on AI comprehension and adoption in a broader sense, including perceptual, semantic, and cognitive computing.

---

[c] CMMI: https://cmmiinstitute.com/cmmi
[d] EKG/MM: https://www.ekgf.org/maturitymodel

## 3.1 AI-Capability Assessment Levels

Each AI-Capability Assessment Level (Table 1) defines the capability requirements that an organisation should fulfil to achieve its AI comprehension and adoption objectives. It describes five levels of development in data management aspects, business readiness, the AI knowledge and skills organisational actors should possess, along with the ethical considerations and risks associated with AI solutions. Based on our review of the literature and experience, the AI-CAM levels are as follows:

**Level 1**: Organisations operating at this level possess basic capabilities to implement entry-level AI solutions. Data governance and management capabilities are immature and poor: Much time and effort are expended on data cleansing prior to analysis. Small-scale pilot projects will be underway; however, they may not be aligned with business strategy goals and objectives. There may also be a lack of exemplars or case studies for decision-makers to draw on. The required knowledge and skills for the comprehension and adoption of AI technologies may not exist. Organisations operating at this level need to begin building or acquiring AI knowledge and skills necessary for desired business use cases related to their business model and competitive strategies. It is also important for organisations to begin evaluating the ethical, regulatory and social responsibilities along with risks—reputational, regulatory, operational risks and so on.

**Level 2**: Organisations operating at this level possess the capabilities to identify the contribution of AI technologies to achieve business objectives. The organisation has an ad hoc approach to data management, governance, architecture, technology and quality on the pilots: Data must be manually cleaned in some cases. An AI-technology pilot is fully deployed in at least one area of the organisation. The ROI for the piloted AI solution is measured to quantify benefits in terms of potential cost reduction, increase in profits or market share. The strategic importance of AI-based solutions is also identified. AI knowledge and skills acquisition by technical staff involved in pilots are at expert levels. The ethical consequences and organisational and business risks of the piloted AI solutions are being carefully evaluated: Data protection measures in pilot areas are managed well.

**Level 3**: Organisations operating at this level will have completed a number of AI-based pilot projects and are brining one or more pilots to scale. The organisation will have achieved and manages the requisite data management capabilities at the enterprise level across all relevant categories. They will be engaging in the quantitative evaluation of a range of AI-based solutions in terms of competitive market share, ROI, cost/benefit, and operational efficiency and effectiveness. Business and IT managers acknowledge that data management capabilities are immature and have developed an integrated strategy for data management, data governance, data architecture, data technology and data quality dimensions. Key organisational staff in strategic and operational business areas are receiving education in AI technologies to enable them to identify, assess and manage opportunities to employ these technologies to operationalise or support the organisation's business model and achieve strategic objectives. Policies and plans to address the ethical issues and organisational risks (including data protection) are also in place.

**Level 4**: Organisations operating at this level will have one or more AI-based applications in operation. Requisite data management capabilities are achieved in the target business use case area(s) and defined for the organisation as a whole. There is a move from siloed structured and unstructured data to a federated and integrated architecture, initially with virtualisation an objective enabled through semantic computing's knowledge representation. AI-based applications' ROI, cost/benefit, operational efficiency, effectiveness, and strategic impact are measured and known. The organisation will be evaluating new business cases, both strategic and operational. Managers and staff in core business areas will have achieved high levels of knowledge and skills on the organisation's existing AI applications. The organisation will have a plan for knowledge and skills acquisition in novel or untried AI technologies. Policies and plans to address the ethical issues and organisational risks (including data protection) are in place and continuously updated in light of novel AI technologies.

**Level 5**: Organisations operating at this level will have several AI-based application(s) in operation across business units and business lines. Organisations are proficient in measuring and assessing the ROI, cost/benefit, operational efficiency, effectiveness, and strategic impact of AI technologies. The organisation continuously evaluates new business cases, both strategic and operational, seeking opportunities to leverage its AI knowledge and skills across these areas. Managers and staff in core business areas will have achieved high levels of knowledge and skills on the organisation's existing AI applications. They will be actively transferring these capabilities to other areas. The organisation will continue to acquire new knowledge and skills in novel or untried AI technologies. Policies to address the ethical issues and organisational risks (including data protection) are continuously reviewed and updated to maintain compliance with existing regulations.

*Table 1: AI-Capability Assessment Levels*

| | AI-CAM Level Criteria |
|---|---|
| **Level 5: Quantitatively Managed** | The organisation is/has: <br>• Socialised and managed its organising vision for AI to provide the framework for interpreting AI, legitimising the adoption and use of AI technologies, and mobilising different communities from IT to business to innovate using it. <br>• Achieved and is managing the requisite data management capabilities at the enterprise level. <br>• Enabling or supporting core business processes using AI-based applications. <br>• Proficient in measuring and assessing the ROI, cost/benefit, operational efficiency and effectiveness and strategic impact of AI technologies. <br>• Identifying additional business cases to apply AI technologies. <br>• Routinely acquiring and transferring new and existing AI-related knowledge and skills throughout the organisation. <br>• Continuously reviewing policies to address the ethical issues and organisational risks to maintain compliance with existing regulations. <br>• Maintaining and managing compliance with data protection regulations. |
| **Level 4: Defined** | The organisation is/has: <br>• Implementing a strategic plan for AI-enabled applications and defining its organising vision. <br>• Operating AI-based application(s) to enable or support a core business process or activity. <br>• Achieved and is managing the requisite data management capabilities in the target business use case area(s). <br>• Measuring and evaluating ROI, cost/benefit, operational efficiency and effectiveness and strategic impact of AI-based technologies. <br>• Evaluating new business cases, both strategic and operational. <br>• Achieving and transferring high levels of knowledge and skills on the organisation's existing AI applications. <br>• Planning for knowledge and skills acquisition in novel or untried AI technologies. <br>• Maintaining and continuously updating policies and plans to address the ethical issues and organisational risks (including data protection). |

| Level 3: Strategic | The organisation is/has:
• Developed a strategic plan and organised vision for AI-enabled applications to digitally transform core business operations and processes.
• Recognised that it does not have the requisite data management capabilities and has developed a strategy for data management, data governance, data architecture, data technology and data quality dimensions.
• Completed several AI-based pilot projects and is bringing one or more pilots to scale.
• Engaging in the quantitative evaluation of a range of AI-based solutions in terms of competitive market share, ROI, cost/benefit, and operational efficiency and effectiveness.
• Capable of evaluating the capabilities of third-party AI technology vendors.
• Educating key organisational staff in strategic and operational business areas in AI technologies to enable them to identify, assess and manage opportunities to employ these technologies to operationalise or support the organisation's business model and achieve strategic objectives.
• Drafting policies and plans to address the ethical issues and organisational risks (including data protection) relating to the strategic use of AI. |
|---|---|
| Level 2: R&D | The organisation is/has:
• Understanding of the business and competitive potential of the AI solutions to informate and automate core business processes.
• An ad hoc approach to data management, governance, architecture, technology and quality on the pilots.
• Implemented business use case pilots for AI solutions in at least one area of the organisation.
• Evaluated AI's potential to increase competitive market share, and the organisation can measure ROI, cost/benefit and operational efficiency and effectiveness.
• Acquiring and building AI knowledge and skills through AI pilots and other sources.
• Evaluated ethical and AI-related risks of pilots, considering the impact on humans, data protection and regulatory compliance. |
| Level 1: Initial | The organisation is/has:
• Poor or basic understanding of the business and competitive potential of the AI solutions.
• Experimented and/or adopted basic AI technologies in, for example, areas such as data analytics.
• Poor data governance across the organisation, which operates a siloed data architecture and where manual data cleansing is the norm.
• Poor alignment of AI-based solutions with business goals and strategic objectives.
• Engaged in knowledge acquisition for understanding and envisioning the potential of AI-based business solutions.
• Considered general ethical issues and risks of AI solutions. |

## 4    AI Capabilities Matrix (AI-CM)

The proposed AI Capabilities Matrix (Table 4) describes the AI knowledge and skills organisational actors should possess at different levels of AI adoption. The matrix aims to provide a general view of the AI knowledge and skills landscape across an organisation and complement the resources that decision-makers can use in their path to AI adoption. The capabilities are matched to types of roles within an organisation, grouped according to three levels of AI proficiency, as defined in Table 2.

Each AI proficiency level builds on the knowledge and skills required at lower levels for each type of organisational actor. For example, the knowledge and skills of chief executive officers (CEOs) and chief technology officers (CTOs), who decide, plan, and drive an AI solution, differ from the knowledge and skills required from those who implement AI solutions (for instance, project leaders, systems administration leaders, system architects, computer and data scientists, knowledge and data engineers, and programmers), as well as from those who participate in other ways in AI projects, such as subject matter experts.

*Table 2: AI Proficiency Levels*

| AI Proficiency Level | Description | Examples of Roles[34-40] |
|---|---|---|
| **Basic** | Basic level of AI proficiency for users within the organisation who interact with AI | Subject-matter expert, trader, administration staff, factory worker, driver, quality control inspector, etc. |
| **Advanced** | Advanced level of AI proficiency for decision-makers who design the organisation's AI strategy | CIO, Head of Compliance, CMO, Head of eCommerce, director, supervisor, senior-management personnel, executive, general manager, etc. |
| **Expert** | Expert level of AI proficiency for technical staff who implement the AI technology | Data scientist, data analyst, machine learning engineer, deep learning engineer, machine learning researcher, deep learning researcher, software engineer, predictive modeller, corporate analytics manager, information strategy manager, computational linguist, computer vision engineer, knowledge engineer, semantic-web consultant, ontologist, ontology manager, applied science manager, ontology expert, knowledge engineering manager, etc. |

*Table 3: AI Capability Assessment Model*

| AI Capability Assessment Model (AI-CAM): Core Areas | | | | | | |
|---|---|---|---|---|---|---|
| Core Areas / Levels | Business | Data | Technology | Organisation | AI Knowledge and Skills[e] | Ethics and Risks |
| **Level 5: Quantitatively Managed** | -An organising vision is socialised and managed to provide the framework for interpreting AI, legitimising the adoption and use of AI technologies, and mobilising different communities from IT to business to innovate using it.<br>-The organisation continuously evaluates, measures, and assesses the ROI, cost/benefit, operational efficiency and effectiveness, and strategic impact of AI technologies.<br>-Additional business use cases are considered in alignment with business model and strategy. | -The following data management (DM) capabilities are fully in place:<br>1. DM Strategy<br>2. DM Business Case and Funding<br>3. DM Program<br>4. Data Governance<br>5. Data Architecture<br>6. Data Technology<br>7. Data Quality Program<br>-Data is identified, defined and governed.<br>-An enterprise data model exists.<br>-Data collections and silos are federated, integrated and virtualised, and completely managed. | -Methodology and metrics for measuring the performance, efficiency and effectiveness of AI technology are developed and applied.<br>-Technological innovation using AI solutions are assessed and managed in line with business metrics.<br>-Technologies from the perceptual, semantic or cognitive computing paradigms are integrated to leverage the full power of AI. | -Senior management make informed decisions at the board level.<br>-Business and technical personnel have the competencies to provide an AI-based solution to internal functions and external clients.<br>-The IT function has internal competencies and external links to acquire or plan, design, develop and deploy novel AI solutions. | -The organisation possesses comprehensive knowledge and skills in all AI computing technologies:<br>1. Perceptual<br>2. Semantic<br>3. Cognitive<br>-AI knowledge and skills are evidenced to varying degrees across the organisation. | -Ethics policies to address the ethical issues and organisational risks to maintain compliance with existing regulations are continuously monitored.<br>-Compliance with data protection regulations is maintained and managed through policies and controls. |
| **Level 4: Defined** | -The organisation has deployed an AI-based application(s) to enable or support a core business process or activity.<br>-It is implementing its strategic plan and has defined its organising vision to deploy AI-enabled solutions in key business processes.<br>-It defines how to measure and evaluate ROI, cost/benefit, operational efficiency and effectiveness, and strategic impact of AI-based technologies.<br>-It is evaluating new business cases, both strategic and operational. | -Data management (DM) capabilities are defined, with a DM Strategy under implementation, as is the DM Business Case and Funding.<br>-The DM Program is also being implemented.<br>-The following capabilities and structures are fully defined for the initial AI deployment, but not across the organisation:<br>• Data Governance<br>• Data Architecture<br>• Data Technology<br>• Data Quality Program | -AI technologies are deployed in production to support a business use case.<br>-Perceptual and semantic computing paradigms are integrated. | -Senior management has defined the need for AI at the board level.<br>-The business team are leveraging the power of AI to inform operational decisions and actions.<br>-The technical team is administering the AI solution in production. | -The business team in the areas where AI is deployed has the requisite capabilities (knowledge and skills) and socialises the defined skillset across the organisation.<br>-The technical team has the required AI knowledge and skills for deploying and managing the selected AI solution. | -Ethics policies to address the ethical issues and organisational risks to maintain compliance with existing regulations are defined.<br>-Compliance with data protection regulations is achieved through policies and controls in target operational areas. |

---

[e] See also the AI Capabilities Matrix (Table 4) for more detailed descriptions.

| AI Capability Assessment Model (AI-CAM): Core Areas | | | | | | |
|---|---|---|---|---|---|---|
| Core Areas / Levels | Business | Data | Technology | Organisation | AI Knowledge and Skills[e] | Ethics and Risks |
| **Level 3: Strategic** | -The organisation has developed a strategic plan and organising vision for AI-enabled applications to digitally transform core business operations and processes.<br>-It has completed several AI-based pilot projects and is bringing one or more pilots to scale in an operational context.<br>-It is engaging in the quantitative assessment of measures to evaluate ROI, cost/benefit, operational efficiency and effectiveness, and strategic impact of AI-based technologies. | -Data management (DM) capabilities are the subject of a DM Strategy, which is under development, as is the DM Business Case and Funding.<br>-The DM Program is also under development, but the following capabilities and structures are not fully defined for the initial AI deployment except for pilot projects:<br>• Data Governance<br>• Data Architecture<br>• Data Technology<br>• Data Quality Program | -An AI solution selected for its strategic impact is implemented and in production in an operational context.<br>-The AI technology solution is based on either the perceptual or semantic paradigm. | -Senior management has the requisite knowledge and skills to develop a business strategy around AI.<br>-The business team are enacting AI's capabilities to inform operational decisions and actions.<br>-The technical team has included business areas to implement the selected AI solution in an Agile project environment. | -The business team in the areas where AI is deployed continue to develop the requisite capabilities (knowledge and skills) from the learnings and outcomes of the pilot projects and applying these in operational contexts.<br>-The technical team continues to develop and hone the required AI skills to develop the selected AI solution or evaluate contractors. | -The strategic impact of ethical issues and various organisational risks are considered.<br>-The implications for compliance with data protection regulations are explored, and policies and controls in target operational areas prepared. |
| **Level 2: R&D** | -The organisation understands the business and competitive potential of AI solutions to informate and automate core business processes.<br>-Business use case pilots for AI solutions have been implemented in at least one area of the organisation. | -Data management (DM) capabilities are recognised as not being sufficient.<br>-The required DM capabilities and structures to enable mature AI deployment are not in place, except in the piloted area. | -The infrastructure for the AI solution implementation is set up.<br>-The technology architecture for developing the AI solution is set up.<br>-Otherwise, if not internally developed, the evaluation of contractors for external development is in place. | -The technical team is created for performing the pilot with the selected AI solution.<br>-The project management methodology is set up. | -The business team in the pilot areas engages in sensemaking on the necessary capabilities (knowledge and skills) required for the pilot's success.<br>-The technical team receives on-the-job training from consultant experts to acquire the required AI skills for completing the pilot. | -The impact of ethical issues and various organisational risks are considered in the context of the pilot.<br>-The implications for compliance with data protection regulations are explored, and policies and controls in target operational areas prepared. |

| AI Capability Assessment Model (AI-CAM): Core Areas | | | | | | |
|---|---|---|---|---|---|---|
| Core Areas / Levels | Business | Data | Technology | Organisation | AI Knowledge and Skills[e] | Ethics and Risks |
| **Level 1: Initial** | -The organisation has a basic understanding of the business and competitive potential of AI solutions.<br>-It has experimented and/or adopted basic AI technologies (e.g., in data analytics).<br>-AI-based [i]solutions are poorly aligned with business goals and strategic objectives.<br>-The organisation is engaged in knowledge acquisition for understanding and envisioning the potential of AI-based business solutions. | -Requirements for data collection infrastructure and data quality are identified for a potential AI solution.<br>- Data management capabilities are poor and require significant development. | -The infrastructure for adopting an AI solution is limited or absent.<br>-Ad hoc development of technical requirements awareness for potential AI solution implementation. | -A business and technology team are created to begin comprehension of AI technologies and business benefits and evaluate the business cases for adopting an AI solution. | -The business team have a rudimentary knowledge of AI and some skills in data analytics.<br>-AI basics/general skills acquisition is underway in the IT function.<br>-Internal and external search is in place to locate the expertise and required AI knowledge based on profiles. | -The strategic impact of ethical issues and various organisational risks are unknown.<br>-The implications for compliance with data protection regulations are unknown and unconsidered. |

The contents of the AI Capabilities Matrix are derived from a synthesis of the capabilities (knowledge and skills) identified in specialist professional reports;[41-45] professional and academic course books, descriptions, and syllabi;[46-49] academic publications;[50-54] governmental webpages;[55,56] job advertisements; and the authors' experience.

The main contributions of this matrix are that: (1) it includes knowledge and skills in all the areas identified in Section 2 as being part of AI technologies (i.e., perceptual, semantic and cognitive computing technologies); and (2) it is cross-functional as its contents are not exclusively concerned with the technical skills for experts or the skills involved in AI adoption at the managerial level, but encompass AI skills specifications for all actors involved in AI adoption across an organisation; and (3) it is linked to an AI capabilities assessment model, the proposed AI-CAM.

The skillsets described in the matrix are not meant to be exhaustive and could indeed be further extended, specified, and linked to specific tasks and tools that would need to be mastered to perform these tasks. However, such specifications would be too context-specific and go beyond this general guide to assist in AI adoption. Note also that the technical skills at the Expert level are not tied to specific roles, such as data scientist or machine learning engineer, as they can (in principle) be acquired irrespective of one's original technical expertise or role within the organisation. Finally, soft skills (e.g., communication, interpersonal skills, time management, independence) and business-level competencies (e.g., understanding business requirements) often required at the Expert level are omitted from the matrix.

The first column, "Capabilities and Technical Areas", lists general competencies (see points A. to E.) with their description. More specific competencies and technical areas relevant to AI are listed under the general competencies (see numbered points). The next three columns specify the skillsets corresponding to each AI proficiency level (Basic, Advanced and Expert). In most cases, it is assumed that the skillsets on the left-side columns also apply to the columns on the right, meaning that someone at the Expert level also possesses the skills at the Basic level. The last column links the skillsets to the AI-CAM levels.

*Table 4: AI Capabilities Matrix*

| Capabilities and Technical Areas | AI Proficiency Level | | | AI-CAM Level (L) |
| --- | --- | --- | --- | --- |
| | **Basic** | **Advanced** | **Expert** | |
| **A. General knowledge of AI** | **Possessing general knowledge of the main questions underlying AI as a discipline and its technologies.** | | | |
| | a. Understanding the differences between Artificial Intelligence and human or animal intelligence.<br>b. Recognising artefacts that use AI and distinguishing them from those that do not.<br>c. Recognising that AI encompasses various areas of activity that make AI technology possible (machine learning, data science, knowledge representation, natural language processing, computer vision, etc.).<br>d. Recognising that developing AI technology involves humans with technical expertise to program and calibrate the AI systems, including supporting technologies. | | | L1:<br>• Advanced<br>• Expert<br><br>L3:<br>• Basic |
| **B. AI Applications** | **Identifying the ways in which AI can be used within the organisation (at individual job and overall organisational levels), its advantages and limits.** | | | |
| 1. AI's strengths and weaknesses | a. Identifying the technologies that are involved in the relevant AI case studies, and their weaknesses and strengths. | a. Identifying tasks and areas within the organisation that could benefit from AI, where human skills are more advantageous, and where human-AI interaction might work best.<br>b. Identifying the AI adoption time and costs for the relevant case studies. | a. Identifying the AI technology and algorithms most suitable for addressing the task.<br>b. Identifying the weaknesses/strengths of the selected AI technologies. | L1:<br>• Advanced<br><br>L3:<br>• Basic<br>• Expert |
| 2. AI capabilities and scope | a. Identifying the AI scenarios that may impact the nature of the work and if this requires training in new areas.<br>b. Understanding that some tasks can be automated with AI, fully or partially, with some human interaction. | a. Understanding what AI can do, to develop business applications with realistic expectations (e.g., an AI chatbot can emulate a human dialogue but with limits).<br>b. Identifying which AI sub-fields apply to the organisation's activities/AI projects: machine learning, natural language processing (NLP), expert systems, computer vision, speech, planning, and robotics. | a. Advising on the technical feasibility of AI business proposals and/or suggesting new ideas.<br>b. Identifying the capabilities of the selected AI technologies. | L1:<br>• Advanced<br><br>L3:<br>• Basic<br>• Expert |

| Capabilities and Technical Areas | AI Proficiency Level | | | AI-CAM Level (L) |
|---|---|---|---|---|
| | Basic | Advanced | Expert | |
| C. Technical skills for AI | Possessing different levels of technical knowledge and skills to participate in the adoption of AI within an organisation. | | | |
| **Perceptual computing** | | | | |
| 1. Machine learning (ML) capabilities | a. Being aware that AI systems are programmed to learn automatically (e.g., to filter spam email, identify images and consumption patterns, etc.). | a. Identifying the different algorithm types and their main usage in practical applications.<br>b. Knowing that ML systems learn from data.<br>c. Explaining ML in non-technical terms.<br>d. Assessing the relevance of ML methods to the organisation's applications. | a. Understanding and implementing ML algorithms: feature engineering, error metrics selection and analysis.<br>b. Using diagnostic tests and interpreting their outputs (e.g., learning curves) to gain insight into the performance and improve it.<br>c. Mastering the mathematical knowledge for ML (e.g., linear algebra, multivariate calculus).<br>d. Applying ML to one or more AI application areas (e.g., NLP, computer vision) and possessing the necessary disciplinary knowledge in the underlying domains (e.g., linguistics, sensors). | L1:<br>• Advanced<br><br>L2:<br>• Basic<br>• Expert |
| **Semantic computing** | | | | |
| 2. Knowledge representation (KR) capabilities | a. Understanding the concept of KR and its main practical usages.<br>b. Identifying how subject matter experts (SMEs) support KR.<br>c. Interacting with technical staff to develop KR models. | a. Knowing the concept of KR and the different KR models (e.g., taxonomies, ontologies, knowledge graphs, etc.).<br>b. Identifying the semantic applications of KR in relevant domains (e.g., health-care, finance, energy, etc.).<br>c. Identifying that KR plays an important role in data harmonisation and integration in an organisation.<br>d. Identifying and ensuring that SMEs fully support the technical staff in KR. | a. Understanding requirements for KR modelling, domain analysis and conceptual analysis.<br>b. Interacting with SMEs to develop KR models.<br>c. Using KR languages (e.g., OWL, RDF, UML), formats (e.g., RDF/XML, Turtle, N-Triples) and query languages (e.g., SPARQL).<br>d. Evaluating NLP techniques for performing KR-related tasks (e.g., ontology learning and population, entity linking, information extraction, etc.).<br>e. Knowing logics and the logical properties of KR languages.<br>f. Using reasoners for consistency checking, classification and deductive inference. | L1:<br>• Advanced<br><br>L2:<br>• Basic<br>• Expert |

| Capabilities and Technical Areas | AI Proficiency Level | | | AI-CAM Level (L) |
|---|---|---|---|---|
| | Basic | Advanced | Expert | |
| **Cognitive computing** | | | | |
| 3. Deep learning (DL) capabilities | a. Being aware that DL is a sub-area of ML.<br>b. Identifying practical applications of DL, for example, in voice recognition (e.g., Siri, Alexa), in autonomous driving (e.g., self-driving cars), etc. | a. Being aware that DL systems are black boxes, which might present risks for given applications (e.g., health-care diagnostic systems).<br>b. Knowing that DL systems learn from data.<br>c. Identifying potential DL-based solutions for the organisation. | a. Understanding and using neural network architectures (e.g., fully connected networks, CNNs, RNNs), methods for training and error analysis.<br>b. Selecting evaluation metrics, and diagnostic testing for interpreting the outputs.<br>c. Possessing the necessary mathematical knowledge.<br>d. Assessing the relevance of DL methods to the organisation's applications.<br>e. Applying DL to one or more AI application areas (e.g., NLP, computer vision) and possessing the necessary disciplinary knowledge in the underlying domains (e.g., linguistics, sensors). | L1:<br>• Advanced<br><br>L2:<br>• Basic<br>• Expert |
| **Supporting Capabilities** | | | | |
| 4. Data engineering capabilities | a. Supporting the data engineers in collecting and annotating the data in case studies.<br>b. Interacting with technical staff to define annotation schemas and annotate data. | a. Knowing that AI systems use data, and the difference between structured and unstructured data.<br>b. Recognising the importance of data quality to train and test AI systems, and of data harmonisation and integration.<br>c. Recognising that data may be biased or contain private information and handling it.<br>d. Ensuring SMEs' support in the data-production processes.<br>e. Identifying that data is a valuable and monetisable asset and brings competitive advantages to the organisation.<br>f. Integrating data protection plans into the AI strategy. | a. Finding or compiling datasets from internal or external sources.<br>b. Evaluating and assessing data sources.<br>c. Exploring, cleaning, transforming, integrating, debiasing, and anonymising data.<br>d. Interacting with SMEs to develop data annotation projects and define annotation schemas.<br>e. Setting up an annotation platform for SMEs.<br>f. Possessing solid scientific foundations and statistical skills.<br>g. Using data management tools with appropriate access and security levels. | L1:<br>• Advanced<br><br>L2:<br>• Basic<br>• Expert |

| Capabilities and Technical Areas | AI Proficiency Level | | | AI-CAM Level (L) |
|---|---|---|---|---|
| | Basic | Advanced | Expert | |
| 5. Software engineering capabilities | | a. Identifying potential software applications, which use AI technologies to facilitate the AI delivery to end-users and customers.<br>b. Including the software applications in the AI strategy.<br>c. Planning the software development capability and maturity models for ensuring quality software applications. | a. Managing the whole development life cycle and models, from requirements elicitation to deployment in production.<br>b. Integrating the AI technologies in the software applications.<br>c. Mastering programming languages for internet, mobile and desktop applications.<br>d. Handling software development methods, both agile and traditional.<br>e. Managing software implementation in different paradigms, such as object-oriented, aspect programming, and internet protocols (e.g., HTML, XML, HTTP request).<br>f. Preparing software documentation. | L1:<br>• Advanced<br><br>L2:<br>• Expert |
| 6. Configuration management (versioning) | | a. Including the configuration management plan in the AI solution for securing the code of the organisation's software. | a. Setting up and managing the configuration management for collaborative team development of the AI solution. | L1:<br>• Advanced<br><br>L2:<br>• Expert |
| 7. Continuous integration | | | a. Setting up and managing a continuous integration platform for delivering the AI solution for deployment in an ambitious frequency (even daily). | L2:<br>• Expert |
| 8. Cloud computing | | a. Identifying and planning a cloud platform for supporting the AI solution development. | a. Selecting a cloud platform for boosting the storage and processing capabilities of the organisation for innovative AI solutions.<br>b. Setting up and managing the AI solution in a cloud platform for providing collaborative development. | L1:<br>• Advanced<br><br>L2:<br>• Expert |
| 9. Virtual server computing | | a. Identifying and evaluating if a virtual server strategy is applicable or acquiring the required hardware for the AI solution. | a. Selecting a virtual server platform vendor or proprietary solution to boost the organisation's storage and processing capabilities for innovative AI solutions.<br>b. Setting up and managing the AI solution in a virtual server configuration. | L1:<br>• Advanced<br><br>L3:<br>• Expert |

| Capabilities and Technical Areas | AI Proficiency Level | | | AI-CAM Level (L) |
|---|---|---|---|---|
| | Basic | Advanced | Expert | |
| 10. System administration | | a. Planning the system administration strategy and staff for administering the AI solution in production. | a. Including the AI solution into the system administration routine tasks for production.<br>b. Knowing the required processing, storage and installation software for setting up the AI solution in production.<br>c. Monitoring the proper functionality of the productive AI solution. | L1:<br>• Advanced<br><br>L3:<br>• Expert |
| **D. Ethical Issues and Risks of AI** | **Identifying societal and ethical implications of AI technology and their potential risks for the organisation.** | | | |
| 1. Capabilities regarding ethical issues and other risks | a. Being aware that AI raises societal and ethical issues.<br>b. Understanding that AI technologies and solutions won't necessarily make an employee redundant.<br>c. Identifying how AI can help them in their daily work. | a. Understanding the societal and ethical questions raised by AI and taking appropriate (business or technical) mitigation measures: privacy issues, surveillance concerns, employment issues (e.g., job destruction), misinformation (e.g., fake news/video generation), singularity/concern about harm to people.<br>b. Considering ethical issues raised by decisions automation with AI, e.g., singularity/concern about harm to people.<br>c. Recognising system development biases resulting from lack of workforce diversity and from biases in the data.<br>d. Recognising risks inherent to black-box systems and the possible need for transparent systems (explainable AI) to ensure accountability. | | L1:<br>• Advanced<br>• Expert<br><br>L3:<br>• Basic |
| | | e. Developing plans to manage ethical risks and data protection. | e. Deploying technical solutions to address ethical issues and mitigate risks. | |

## 5 Discussion and Conclusions

This paper's contribution is a 5-level AI Capability Assessment Model (AI-CAM) and a related AI Capabilities Matrix (AI-CM) which are tools aimed at assisting organisations in AI comprehension and adoption. These tools were developed with executives and leading stakeholders in mind. They are based on a wider definition of AI than usually accounted for in AI adoption models and thus include requirements not only regarding data governance and management issues, and the use of machine learning technologies, but also semantic technologies based on knowledge representation resources. These models encompass technologies for emulating human-like reasoning. Therefore, these tools are particularly suited for seeking to automate tasks involving, for example, querying and reasoning over very large amounts of textual data.

The AI-CAM covers the core areas to be developed to achieve different maturity levels in AI capabilities: business, data, technology, organisation, AI skills, business and operational risks, and ethical questions. Therefore, the AI-CAM provides a valuable resource to organisations to help assess organisational AI capability maturity and determine if it can achieve its strategic and operational objectives using AI. The model specifies the required capabilities required to achieve desired organisational outcomes. The higher the level of capabilities demonstrated across the various dimensions, the greater the potential for success in AI adoption and innovation. An organisation can therefore determine if a capabilities gap exists and use the model and the related AI-CM to close the gap in line with achieving the perceived strategic and operational benefits of AI adoption.

The AI-CM addresses the knowledge and skills needed to reach each level of the AI-CAM. The AI-CM aims at filling the functional, disciplinary and modelling gaps identified in our literature review by (1) specifying AI capabilities and skills at different levels of AI proficiency, (2) in all disciplinary and technical areas covered by AI adoption, including KR, and by (3) linking the capabilities and skills to different stages of AI adoption within an organisation.

It is inspired by the *AI literacy* framework,[57] which identifies the general AI capabilities that everyone ought to possess. The authors of this framework "define *AI literacy* as *a set of competencies that enables individuals to critically evaluate AI technologies; communicate and collaborate effectively with AI; and use AI as a tool online, at home, and in the workplace*".[58] It thus fits the whole range of professionals involved in an organisation's AI adoption path. It contributes to existing proposals by introducing knowledge and skills requirements at all levels of an organisation categorised into three levels of AI proficiency: Basic level for users who interact with AI and participate in AI adoption, Advanced level for decision-makers who design AI strategies, and Expert level for technical staff who implement the AI technology.

While seeking to include a wide range of core skills based on recent professional and academic literature, the AI Capabilities Matrix is by no means a systematic and exhaustive inventory of relevant skills. It is meant to be extended, refined and regularly updated with more systematic reviews of sources in the areas concerned. These tools would also require empirical validation, which we leave for future work.

In conclusion, the AI-CAM and AI-CM offer an open-source resource for organisations to help maximise the benefits of AI adoption and use. The models may also be used to assess the capabilities of AI vendors, to identify potential AI-washing and close the gap between the promise and reality of AI in organisations.

*Acknowledgements*

This work was partially supported by Enterprise Ireland and the Marie Skłodowska-Curie Actions through the Career-FIT Fellowships MF20180003 and MF20180009. Career-FIT has received funding from the

European Union's Horizon2020 research and innovation programme under the Marie Skłodowska-Curie grant agreement No. 713654.

---